\documentclass[iop]{emulateapj}

\usepackage{natbib, graphicx, times, amsmath, epsfig, amssymb}
\usepackage{amsmath}
\usepackage{verbatim}
\def\simgt{\lower.5ex\hbox{\gtsima}} 
\def\gtsima{$\; \buildrel > \over \sim \;$} 

\begin{document}

\submitted{Accepted for publication in ApJ Letters on 4 November 2013}

\title{Detectability of Free Floating Planets in Open Clusters with JWST}
\author{Fabio Pacucci$^1$, Andrea Ferrara$^1$, Elena D'Onghia$^{2,3}$ \\
$^1$Scuola Normale Superiore, Piazza dei Cavalieri, 7  56126 Pisa (Italy) \\
$^2$University of Wisconsin, 475 Charter St., Madison, WI 53706 (USA) \\
$^3$Alfred P. Sloan Fellow}

\begin{abstract}
Recent observations have shown the presence of extra-solar planets in Galactic open stellar clusters, as in the Praesepe (M44). These systems provide a favorable environment for planetary formation due to the high heavy-element content exhibited by the majority of their population. The large stellar density, and corresponding high close-encounter event rate, may induce strong perturbations of planetary orbits with large semimajor axes. Here we present a set of $N$-body simulations implementing a novel scheme to treat the tidal effects of external stellar perturbers on planetary orbit eccentricity and inclination. By simulating five nearby open clusters we determine the rate of occurrence of bodies extracted from their parent stellar system by quasi-impulsive tidal interactions. We find that the specific free-floating planet production rate $\dot N_o$ (total number of free-floating planets per unit of time, normalized by the total number of stars) is proportional to the stellar density $\rho_\star$ of the cluster: $\dot N_o = \alpha \rho_\star$, with $\alpha = (23 \pm 5)\times10^{-6} {\rm pc^3}\,{\rm Myr}^{-1}$. For the Pleiades (M45) we predict that $\sim 26\%$ of stars should have lost their planets. This raises the exciting possibility of directly observing these wandering planets with the James Webb Space Telescope in the NIR band. Assuming a surface temperature of the planet of $\sim 500 $ K, a free-floating planet of Jupiter size inside the Pleiades would have a specific flux of $F_\nu$ (4.4 $\mu$m) $\approx 4\times10^2$ nJy, which would lead to a very clear detection ($S/N \sim 100$) in only one hour of integration.
\end{abstract}


\section{Introduction}
\label{sec:intro}
The first detection of an extra-solar planet orbiting a Sun-like star in a dense stellar environment
dates back to 2012, when a NASA-funded team discovered Pr0201b and Pr0211b inside the Praesepe open cluster \citep{Quinn2012}. This cluster contains more than 1000 stars, having a core radius of $\approx1.3$ pc and a distance from the Earth of $\approx600$ pc \citep{Kraus2007}. The newly discovered planet belongs to the type of extra-solar planets termed  Hot Jupiters (\citealt{Mayor1995, Charbonneau2000}), i.e., massive gas giants that, unlike Jupiter, orbit very close to their parent stars. Hot Jupiters are the easiest planets to be detected with the radial velocity or transit methods, due to their high mass and small semimajor axis \citep{Moulds2013}. However, the origin of this extra-solar planet population is still uncertain.  Several studies suggested  that it is likely to result from a  planetary migration process that dramatically decreases the semimajor axis of giant  planets formed far away from the star, beyond the so-called ice line (\citealt{Lin1976, Goldreich1980, Fabrycky2007}).

 Given the first confirmed planet detection in the Praesepe open cluster, it is likely that several other planets are also present in high-density environments. In particular, a study by \cite{Meibom2013} shows that the frequency of planets inside of clusters is the same as in the field. Indeed,  the high metallicity found in most of the open clusters facilitates the process of planetary formation (\citealt{Fischer2005,Chatterjee2012}) and their relatively high stellar density, about 500 times larger than in the Solar neighborhood \citep{Nilakshi2002}, makes stellar fly-by's  a very likely process. As a result, close encounters between stars in an open cluster are common and may dramatically affect the orbits of planets with large semimajor axes  \citep{Adams2006}. As planetary eccentricities grow, they may get captured into closer orbits, becoming Hot Jupiters; see, e.g., \cite{Wu2012}.

Stellar fly-by's and their effects on planets have been recently studied in various contexts. In the Solar System, for instance, it is thought that the orbit of Sedna is a consequence of stellar fly-by's \citep{Brasser2012}, and recently \cite{Adams2010} discussed the truncation of the Kuiper Belt by fly-by's. In addition, stellar fly-by's have also been investigated in more recent works as a cause of planetary ejections (\citealt{Veras2012, Ovelar2012, Craig2013, Parker2012}). Most of the theories of planet formation suggest that planets should reside in resonances \citep{Masset2001}, but these are found to be seldom \citep{Wright2011b}.  One possibility is that stellar fly-by perturbations  lead to the break-up of multiresonant states (\citealt{Morbidelli2007, Batygin2010}), triggering large-scale instabilities  that could explain the random $a$-$e$ distribution of exoplanets. Additional complications arise from the fact that solar systems can eject planets on their own through instabilities (\citealt{Rasio1996, Nesvorny2011, Batygin2012, Nesvorny2012, Boley2012}).

In what follows we present the results of numerical simulations following the impulsive perturbation of planetary orbits with orbital distances from the parent star in the range $(5-60) \, AU$, roughly from the orbital distance of Jupiter to the farthest extra-solar planets with a clear semimajor axis detection discovered so far (see exoplanets.org, \citealt{Wright2011}). Several planets with much
larger semimajor axes have been discovered \citep{Luhman2011}, but the associated uncertainty is also much larger. Our aim is to show that a sizable fraction of planets within this distance range are ejected from their planetary systems by tidal interactions with nearby stars.
 
 \section{METHODOLOGY}
\label{sec:simulation}
 We have developed a numerical code to accurately describe the tidal effects of a gravitational encounter between a Jupiter-like planet orbiting a central
 star, and a perturbing star moving on a straight line or parabolic trajectory (see \citealt{Donghia2010} for the geometry of the problem). The close encounter has been studied in a four-parameter space, namely
 (a) the distance between the parent star and the planet, $d_p$, 
 (b) the interstellar distance at pericenter, $D$, 
 (c) the inclination between the perturber and planet orbital planes, $i$, 
 and (d) the relative velocity of the perturber, $v_{\rm rel}$. Each parameter interval has been sampled with five points, for a total of 625 different values. 
 For any given point  in the parameter space, the resulting variations of the orbital eccentricity and inclination are 
 saved in the ``interaction matrix". 
 This matrix is used as a sub-grid model for a $N$-body simulation of an open cluster. The use of a sub-grid approach presents some limitations with respect to a fully self-consistent model of $2N$ particles. Nonetheless, the sub-grid model is less computationally expensive than other methods, allowing us to build a better statistics on the escaping planets.
 
  \subsection{Building the Interaction Matrix}
The following Table 1 summarizes the values of the interaction parameters:
\begin{center}
    \textbf{Table 1} \\*
    Interaction Parameters
    
    \begin{tabular}{c c c c}
    \hline
    \hline
    $D$ (AU) & $v_{\rm rel}$ (AU/yr) & $i$ ($^{\circ}$) & $d_p$ (AU) \\ \hline
    30 & 0.01 & 0 & 5 \\ \hline
    100 & 0.05 & 10 & 15 \\ \hline
    1000 & 0.1 & 30 & 30  \\ \hline
    10000 & 0.5 & 45 & 45 \\ \hline
    100000 & 1.0 & 60 & 60 \\ \hline
    \end{tabular}
\end{center}
A gravitational encounter between two stars yields a modification of the planetary orbit, caused by an energy transfer between the perturbing star and the planet. The interaction matrix provides information about the values of $\Delta e$ and $\Delta i$ for all the evaluation points.

Denoting by $M$ the mass of both stars and $m$ the mass of the planet, we fix the system of reference on the parent star and suppose that the perturber follows a straight line trajectory. The parabolic trajectory case has also been implemented as in \cite{Donghia2010}, but the total number of free-floating planets differs in the two cases by less than $1.5\%$ for all the five clusters. The initial orbital plane of the planet coincides with the $x-y$ plane, with the $x$ axis pointing in the direction of the pericenter. 
In this reference frame, the force acting on the planet can be decomposed into the centripetal force exerted from the parent star and the tidal force on the planet, caused by the passage of the perturber. The equations of motion are solved via a Runge-Kutta fourth-order method. The orbital eccentricity and the inclination are evaluated during the integration with the following expressions:
\begin{equation}
i = \arctan \left(\frac{z}{\sqrt{x^2+y^2}} \right)
\end{equation}
\begin{equation}
e = \sqrt{\left(\frac{1+2\epsilon{\lambda^2}}{\mu^2} \right)}
\end{equation}
where $\epsilon= E/m$ is the specific orbital energy, $\lambda=L/m$ is the specific angular momentum and $\mu= G(M+m)$.
The parent star is displaced at the origin of the reference frame and remains at rest during the entire evolution of the system. A single planet is assigned to each star and it is displaced on a circular orbit at a distance specified by the interaction matrix.
The perturbing star has the following initial conditions for the position:
\begin{equation}
[x, y, z] = [D \cos(i), y_{\rm shift}, D \sin(i)]
\end{equation}
The value of $y_{\rm shift}$ is calculated assuming that the perturber reaches the pericenter at half of the integration time $t_{\rm end}$.
We choose to integrate up to a time when the ratio between the magnitude of the perturber-planet force and the magnitude of the parent star-planet force is equal to a constant, $T_{\rm int}$:
\begin{equation}
t_{\rm end}=\frac{1}{v_0}\times \sqrt{ \frac{M d_p^2}{mT_{\rm int}}-(D-d_p)^2 }
\end{equation}
The time evolution (for a single point in the parameter space) yields the total variations in eccentricity and inclination.
 
 \subsection{The Open Cluster Simulation}
 The code NBODY6 \citep{NBODY6} has been modified, with the addition of a separate routine 
 to manage the planetary configurations, which is called once at each time step.  A simulation of $N_{\rm stars}$ stars 
 has been initialized with a Plummer steady-state distribution \citep{Ernst2011} in the phase-space. All these objects are one solar mass stars, without any stellar evolution, and the cluster is supposed to be isolated in space. This physical system has been simulated up to the age $\tau$ of the stellar cluster. 
 Our simple model consists of only single stars, although the binary star fractions in open clusters are observed to be high. This is a limitation of our model since the encounters with binary stars would likely be quite common, but we do believe that our simple model provides a robust lower limit on the number of free-floating planets in these high-density stellar environment. 

The choice of the Plummer model instead of more accurate models (e. g. the King ones) is justified by the fact that, unlike the  globular clusters,  open clusters are not completely relaxed objects, so that the difference between the two models is not significant.
The cluster is subject to a natural evaporation with time due to high-speed encounters which eject stars from the cluster. As a consequence, the number of interacting stars is also variable with time.

Most of the empirical radial distributions of exoplanets found in literature are related to inner planets, with semimajor axis smaller than a few AUs (see e.g. \citealt{Boivard2013}). Unfortunately, planets in exactly the range we are primarily interested in $(\sim 10\, AU)$ are the most difficult to detect. Given this observational bias and focusing on the range $(5 - 60)\, AU$ of semimajor axis, it is possible to fit the observed distribution of planets (retrieved from the database exoplanets.org) with an exponential decay function, namely $N(r) = A \times e^{(-r/a)}$, where $A$ is a normalization constant, $a$ is the e-folding length and $N(r)$ is the number of planets with a semi-major axis distance r from the central star. The initial distribution of eccentricities is flat, with $e_0 = 0.042$.
 
 Every gravitational encounter between a pair of stars causes a tidally-driven modification of the planetary orbit of their two planets.
 For every star in the simulation, the routine loops over all the other stars and computes $D$, $i$ and $v_{\rm rel}$, while $d_p$ has already been assigned.
The corresponding variations of the planetary inclination and eccentricities are calculated by interpolation from the sub-grid model and applied only when the interstellar distance calculated between the pair of stars has increased with respect to the same quantity calculated at the last time step. This approximation is supported by the fact that the variations of eccentricities and inclinations in the tested encounters are very focused around the time of pericenter. The accuracy in the determination of the pericenter is related to the magnitude of the time step used by NBODY6.
If the pericenter falls between two time steps, the values calculated from the interaction matrix for the orbital modifications are always underestimated. 
A partial solution for this problem is to increase the values of $\Delta{e}$ and $\Delta{i}$ by a fraction of their values which is proportional to the following quantity:
\begin{equation}
\frac{Rv_{\rm rel}\Delta{T}}{2D}
\end{equation}
where $\Delta{T}$ is the numerical time step (when it goes to zero, the correction vanishes) and $R$ is a random number with uniform probability between 0 and 1, which expresses our lack of knowledge of the real position of the pericenter.
We suppose that the orbit of each planet remains circular throughout the simulation. Firstly, we assumed that the radius remains constant: this situation is physically unrealistic, but offers a very secure lower-limit for the number of free-floating planets. Then, we assumed that the radius is modified proportionally to the eccentricity, with the simple scaling law: $d_{new} \sim d_{old} \times (1+e)$. In this second case, the total number of free-floating planets increases by roughly 9\%.

Several checks on the overall consistency of the simulation have been performed, as the one concerning the effect of the mean field, i.e. the cumulative gravitational effects of the stars that are very far from the star under investigation. Its effect on the orbital parameters of a planet is negligible when compared to the gravitational effects of a strong encounter.
We perform a fit of the ever-increasing eccentricity of the planet before a strong encounter and extrapolate the fit up to the final time of the simulation, noticing that the corresponding variation of the eccentricity ($\Delta{e}\sim10^{-4}$) is negligible for our purposes. For the very same reason, also the $N$-body encounters, with $N>2$, have negligible effects on the extraction rate of the planets.
 
 If the total energy of a planet becomes positive, it escapes from its planetary system. 
 Such free-floating planet is treated as a separate particle, with velocity magnitude equal to the escape velocity and random direction. Wandering planets interact with all the other stars, but not with other wandering planets and their dynamics is followed by another routine added to NBODY6.
 If a free-floating planet approaches another star within a distance of 250 AU, its total energy, $E$, with respect to that star is computed and, if negative, the planet is considered 
 as re-captured and is eliminated from the list of free-floating planets \citep{Perets2012}.
 
\section{Results} 
A summary of our results for five different galactic open clusters (NGC188, NGC6530, M16, M44, M45) is reported in Figure 1, 
where we plot the specific free-floating planet production rate (total number of free-floating planets per unit of time, normalized by the total number of stars) 
as a function of the central stellar density of the open cluster.

\begin{figure}
\includegraphics[width=0.5\textwidth]{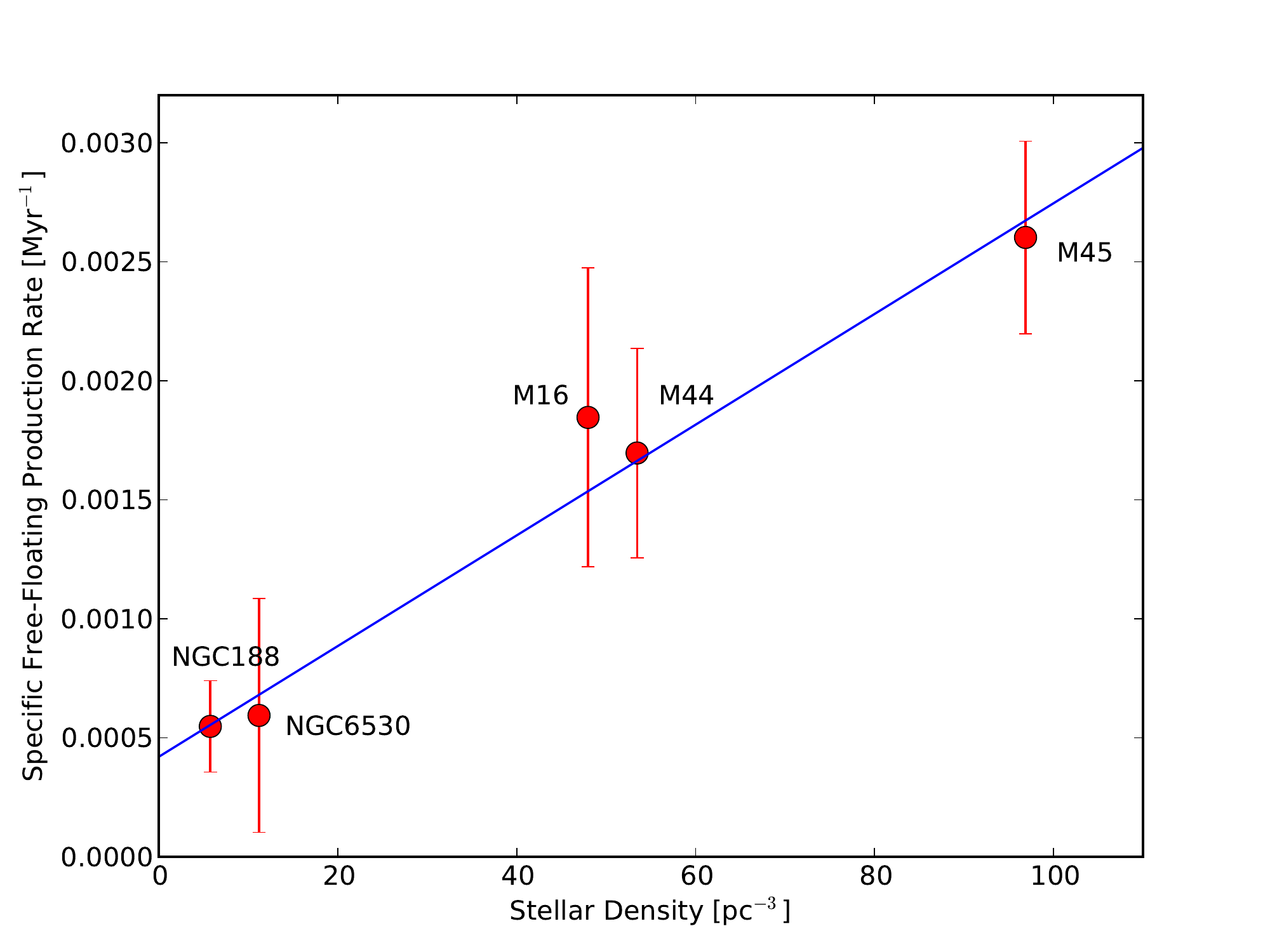}
\caption{Specific free-floating planet production rate as a function of the open cluster central stellar density, for five different galactic open clusters.}
\label{fig:trend}
\end{figure}

For our simulations, we have used the parameters shown in Table 2 for the five galactic open stellar clusters:
\begin{center}
    \textbf{Table 2} \\*
    Open Clusters Data

    \begin{tabular}{c c c c}
    \hline
    \hline
     & Number of stars & Core radius (pc) & Age (Myr) \\ \hline
    M44 & 1391 & 1.3 & 700 \\ \hline
    M45 & 3151 & 1.4 & 100 \\ \hline
    M16 & 1560 & 1.4 & 2.5  \\ \hline
    NGC6530 & 365 & 1.4 &  2 \\ \hline
    NGC188 & 5388 & 2.9 & 6000 \\ \hline
    \end{tabular}
\end{center}
The number of stars is derived using the astronomical database SIMBAD, which also allows to set the membership probability.
In any case, the five different galactic open clusters present in the paper have been chosen in order to provide a wide range of stellar densities in the cluster, as long as different ages.

The relation between the specific free-floating planet production rate $\dot N_o$ 
and the stellar density $\rho_\star$ is given by $\dot N_o = \alpha \rho_\star$ where $\alpha = (23 \pm 5)\times10^{-6} {\rm pc^3}\,{\rm Myr}^{-1}$. See also \cite{Spurzem2009} where similar results are obtained.
This strikingly simple relation indicates that the presence and abundance of free-floating planets is clearly related to the environmental stellar density.
In addition, these clusters formed more massive than they are today and with very different central densities and density structures, so that our predictions for the free-floating planets formation rate are again strictly lower limit estimates.
The evolution of the eccentricity of the planetary population of the Pleiades is shown in Figure 2. 

\begin{figure}
\includegraphics[width=0.5\textwidth]{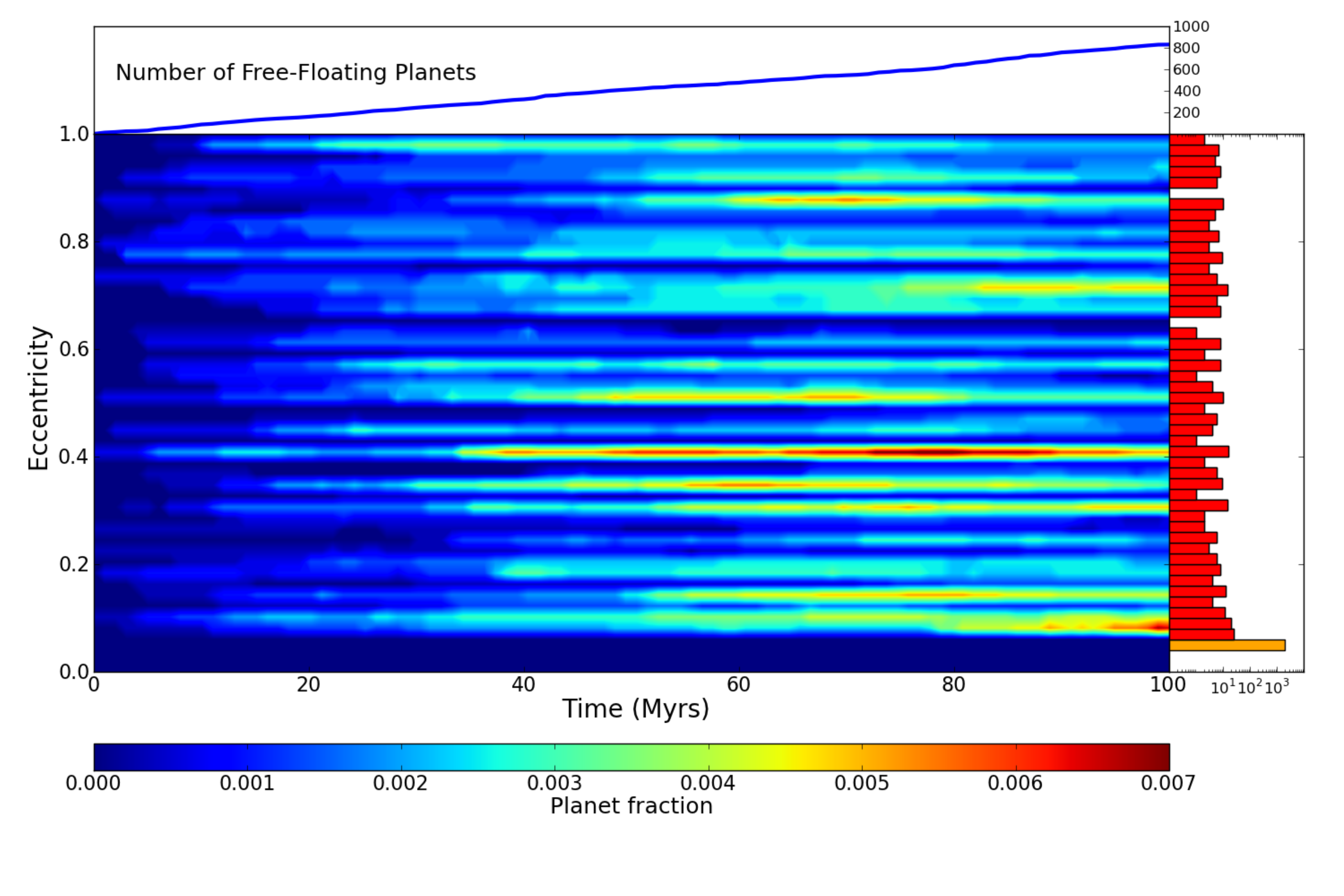}
\caption{Time evolution of the eccentricities of the planetary population of the Pleiades, up to 100 Myrs of evolution. The contour plot shows the distribution of eccentricities as a function of time, along with the distribution of eccentricities for the population of bounded planets at the final simulation time. The lowermost orange bar refer to (unperturbed) planets that are still in the initial eccentricity bin ($e_0 \approx 0.042$). This population is the dominant one and it has not been included in the contour plot for the sake of clarity. Finally, the top panel reports the cumulative number of free-floating planets as a function of time, reaching a level $\sim 800$ at the end of the simulation.}
\label{fig:eccentricity}
\end{figure}

The contour plot describes the overall increase with time of the eccentricity of the planetary population. A direct comparison between our results, obtained with a sub-grid approach, and a fully self-consistent integration of a system of $2N$ particles has been made. In the case of the Pleiades cluster, the total number of escapers is higher by less than 10\% with the use of a self-consistent integration.

\section{Observability of Free-Floating Planets}
 As free-floating planets do not orbit around a star, their direct detection, usually hampered by the unmanageably high contrast with the star for normal planets (\citealt{Sumi2011, Dong2013, Delorme2012, Beichman2013, Gould2013}), 
 might be possible in the infrared with JWST \citep{Burrows2003}. Figure 3 shows a simulated image (for the typical 2.2 arcmin field of view of the JWST/NIRCam instrument) 
 of the central region of the Pleiades cluster, evolved up to 10 Myrs. 
 
\begin{figure}
\includegraphics[width=0.5\textwidth]{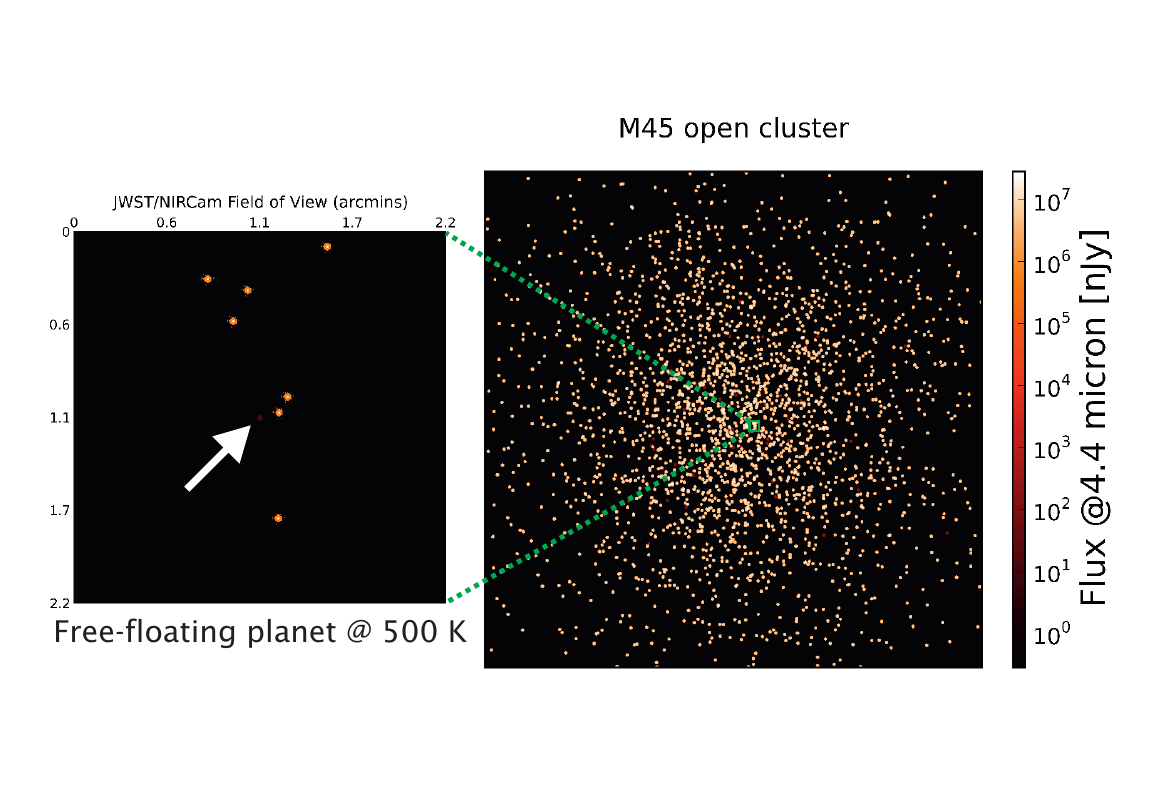}
\caption{Right panel: Simulated image of the central region of the Pleiades, after 10 Myrs of evolution showing the expected flux at $\lambda=4.4 \, \mu m$ from stars and free-floating planets. Left panel: a typical field of the NIRCam camera onboard the future JWST observatory. The arrow shows the location of a free-floating planet with a surface temperature of $\sim 500$ K.}
\label{fig:simulation}
\end{figure}

 The values of the flux (in nJy, at 4.4 $\mu$m) are calculated assuming the appropriate 
 spectrum for both the stars (assumed to be Sun-like) and the free-floating planets (assumed to have a surface temperature $T_{\rm eff} = 500$ K). 
 The free-floating planet inside the field is indicated with a white arrow; its flux is $F_\nu$ (4.4 $\mu$m) $\approx 4\times10^2$ nJy and it is detectable in $3600$ s of integration time with a $S/N\sim100$ (see JWST Exposure Time Calculator for reference).
 
 Free-floating planets can then be detected by the NIRCam if $T_{eff} \simgt 300$ K, because for lower surface temperatures the specific flux would be too faint. For example, with $T_s = 300$ K, the corresponding specific flux would be $F_\nu (4.4 \mu m) \approx 8$ nJy, which needs $\sim 1$ hr of integration to reach $S/N \sim 2$ level.
 
 The actual observation of free-floating planets implies the possibility of distinguishing them from very far away stars (which are redshifted to the IR) and brown dwarfs already formed in isolation inside the open cluster.
The fact that background stars are dynamically detached should serve as a mean to distinguish them from our free-floating planets population, thanks to their higher redshift and possibly different proper motion.
The distinction between free-floating planets extracted from bounded planetary systems and isolated brown dwarfs is more difficult. The rate of formation of isolated brown dwarfs inside open clusters may be deduced from the low-mass end of the IMF. Comparing the theoretical prediction with the observed population of free-floating planets, the excess probability over the formation rate of isolated brown dwarfs could be accounted as a measure of the occurrence of the free-floating planets, predicted by our model.

\section{Conclusions}
 In this work we have shown that tidal interactions between stars in relatively high-density environments, such as open stellar clusters, may well cause the ejection of external planets from their systems. By means of using an accurate sub-grid model setup for a large parameter space, we have performed a purely $N$-body simulation of different Galactic open stellar clusters, with different morphological properties and ages. Assigning to each star a planetary body of Jupiter-mass and variable distance from the parent star, we have studied the occurrence of free-floating planets over the entire evolution of each cluster, also accounting for the possibility of a planet being recaptured by another star. In particular, we have shown that the specific free-floating planet production rate $\dot N_o$ is linearly related with the stellar density of the cluster $\rho_\star$: $\dot N_o = \alpha \rho_\star$, with $\alpha = (23 \pm 5)\times10^{-6} {\rm pc^3}\,{\rm Myr}^{-1}$.  Specifically, for the Pleiades we predict that $\sim 26\%$ of the stars should have lost their planet during the evolution of the cluster.

The high contrast with the star for bounded planets poses very strict, and usually unmanageable, limits on their direct observation. On the contrary, the large population of free-floating planets predicted in this work might be detected with the next generation of space telescopes, given that their surface temperature is sufficiently high. The study on observability shows that even with a relatively low-temperature planet at $300$ K, a detection may be feasible in the infrared band, using the NIRcam instrument onboard the future JWST observatory. A clear detection becomes much more feasible if the free-floating planets (or at least a fraction of them) reach temperatures of at least $\sim 500$ K. The coolest brown dwarf candidates have very similar surface temperatures \citep{Luhman2012}, so the lower bound of  $\sim 300$ K might be reached by a large fraction of the population of free-floating planets.
The detection of free-floating planets would open a new pathway to exoplanetary and stellar-cluster studies, allowing us to test the survival and evolution of planets under extreme interstellar conditions.

\vspace{10 pt}

We are grateful to Kostantin Batygin for suggestions and a careful reading of the manuscript. ED gratefully acknowledges the support of the Alfred P. Sloan Foundation.

\bibliographystyle{apj}


\end{document}